# Tunable magnetic and transport properties of $Mn_3Ga$ thin films on Ta/Ru seedlayer

Fang Hu[1], Guizhou Xu[1,a)], Yurong You[1], Zhi Zhang[1], Zhan Xu[1], Yuanyuan Gong[1], Er Liu[1], Hongguo Zhang[2], Enke Liu[3], Wenhong Wang[3], Feng Xu[1,a)]

1 School of Materials Science and Engineering, Nanjing University of Science and Technology, Nanjing 210094, China
2 College of Materials Science and Engineering, Beijing University of Technology, Beijing 100124, China
3 State Key Laboratory for Magnetism, Beijing National Laboratory for Condensed Matter Physics, Institute of Physics, Chinese Academy of Sciences, Beijing 100190, China

**Abstract**

Hexagonal $D0_{19}$-type $Mn_3Z$ alloys that possess large anomalous and topological-like Hall effects have attracted much attention due to their great potential in the antiferromagnetic spintronic devices. Here, we report the preparation of $Mn_3Ga$ film in both tetragonal and hexagonal phases with a tuned Ta/Ru seed layer on the thermally oxidized Si substrate. A large coercivity together with a large anomalous Hall resistivity is found in the Ta-only sample with mixed tetragonal phase. By increasing the thickness of Ru layer, the tetragonal phase gradually disappears and a relatively pure hexagonal phase is obtained in the Ta(5)/Ru(30) buffered sample. Further magnetic and transport measurements revealed that the anomalous Hall conductivity nearly vanishes in the pure hexagonal sample, while an abnormal asymmetric hump structure emerges in the low field region. The extracted additional Hall term is robust in a large temperature range and presents a sign reversal above 200K. The abnormal Hall properties are proposed to be closely related with the frustrated spin structure of $D0_{19}$ $Mn_3Ga$.

---

a) Authors to whom correspondence should be addressed. Electronic mail: gzxu@njust.edu.cn, xufeng@njust.edu.cn

## I. INTRODUCTION

Recently, magnetic alloys with frustrated kagome lattice have received much attention due to their intriguing physical properties and great potential in the antiferromagnetic (AFM) spintronic devices[1-3]. For instance, in the non-collinear antiferromagnets (AFM) of $D0_{19}$-type $Mn_3Sn$ and $Mn_3Ge$ single-crystals[4,5], due to the existence of a Mn-kagome lattice, an exceptionally large anomalous Hall effect (AHE) has been identified. It is also predicted that the associated non-vanishing Berry phase in these alloys can generate large spin Hall effects[6] and orbital ferromagnetism[7], where the former has already been verified in $Mn_3Ir$ thin film[8]. Very recently, a topological-like Hall effect below a critical temperature has been reported in the hexagonal $Mn_3Ga$ polycrystals[9]. For $Mn_3Ga$, abundant researches have been carried out on the $D0_{22}$ tetragonal phase[10-13], while studies on the hexagonal one are still lacking. As thin films can generally enhance the stability of specific domain formation[14,15] and are indispensable for application in spintronic devices, here we intend to study $Mn_3Ga$ in the form of thin films with various phases and concentrate on the magnetic and transport properties of hexagonal $Mn_3Ga$ phase.

$Mn_3Ga$ commonly exhibits two stable phases, hexagonal $D0_{19}$ and tetragonal $D0_{22}$, depending on the history of thermal treatment. The former structure can be obtained by a high temperature annealing (~600–800°C) followed with quenching[9,16], while the latter one can be obtained by low temperature annealing (350–450°C)[10,11,16]. It is the result of a hexagonal-tetragonal phase transformation that occurs at about 770 K. It is also reported recently that, with a rapid quenching method like melt-spinning, $Mn_3Ga$ can present a metastable face-centered-cubic structure[17,18]. The $D0_{22}$ tetragonal $Mn_3Ga$ usually exhibits large perpendicular magnetic crystalline anisotropy[12], along with a small saturation magnetic moment, a low Gilbert damping coefficient and a large spin polarization. It thus has been recommended for application in spin transfer torque devices[10,11,12].

Due to the sensitivity of structure, the state of the $Mn_3Ga$ film can be strongly affected by the deposition temperature and choice of seed layer[19]. Tetragonal $Mn_{3-x}Ga$ films have been successfully grown on MgO or Si substrates with various seed layers, under proper heat treatments[12,13,20]. By choosing special seed layer like $Pt(111)$[21] and $Ru(0001)$[22] or annealing at a higher temperature[23], the hexagonal $D0_{19}$ phase can also be obtained. In this work, we successfully tuned the structure of $Mn_3Ga$ thin films by magnetron sputtering on Ta/Ru-buffered thermally oxidized Si substrate. Magnetic and Hall properties varied distinctively for different phases, and a topological-like Hall effect was observed in the Ta(5)/Ru(30)-buffered $Mn_3Ga$ film.

## II. EXPERIMENTAL METHODS

All film layers were deposited by RF-magnetron sputtering with a base pressure of $\sim 5 \times 10^{-5}$ Pa. The RF power of 80W was applied, under the sputtering pressure of ~1Pa. The deposition rate for Mn$_3$Ga layer is 10nm/min. The thickness of all the Mn$_3$Ga films is 200nm. We used a stoichiometric Mn$_3$Ga target, but the film composition scanned by the energy-dispersive X-ray spectroscopy (EDX) is about 3.4:1 for Mn:Ga. According to the recently detected Mn-Ga binary phase diagram, the hexagonal D0$_{19}$ phase can cover a wider composition range that contains the above value[24]. A 5nm Ta under-layer was pre-sputtered on thermally oxidized Si substrate. Ru and Mn$_3$Ga layer were subsequently deposited at room temperature. The stack of the multilayer film is shown in Fig. 1(a). The whole wafer was then in-situ annealed at 550°C for 20 minutes. The structure of the films was investigated by X-ray diffraction (XRD, Cu Kα) and transmission electron microscopy (TEM, FEI Tecnai 20). The surface morphology of the samples was scanned by atomic force microscopy (AFM, Bruker Multimode 8). The magnetic and transport properties were measured in a multifunctional physics property measurement system (PPMS, Quantum Design).

## III. RESULTS AND DISCUSSION

We chose the Mn$_3$Ga samples with no Ru layer (designated as Ta-only sample afterwards) and Ru=10, 20, 30nm to investigate. The surfaces of all the films are rather flat and smooth. One typical example of Ru-30nm is shown in Fig. 1(b), with the root-mean-square (rms) surface roughness to be 1.57nm. The slow scanned X-ray diffraction pattern of the films is revealed in Fig.1(d). In the Ta-only sample, there are two split peaks centered at about 40.5° and 40.9°, and one at 42.4°. While for the Ru-buffered films, the spilt peaks disappeared, with a single peak left at around 40.9°. This indicates that Ta-only sample should contain two phases of Mn$_3$Ga, and one of them will gradually disappear with increasing Ru thickness. We attribute the split peaks around 40.9° to the coexistence of the tetragonal D0$_{22}$ and hexagonal D0$_{19}$ phases, as the (112) peak of D0$_{22}$ phase and (002) peak of D0$_{19}$ Mn$_3$Ga are rather close in bulk[17,24,25]. The subsequent magnetization measurements also confirm the co-existence of a hard ferromagnetic and AFM phase in the Ta-only sample. With insertion of the Ru layer, an additional peak at ~42.1° emerges, which is indexed as the hexagonal Ru(0001) direction. This peak is near to the (201) peak of D0$_{19}$ Mn$_3$Ga[18,25]. From the peaks at 40.9° and 42.4° in the Ta-only sample, the lattice constant for the hexagonal one is determined to be about a=5.62Å and c=4.41Å. From the Tetra-(112) peak, we determine the lattice constant for the tetragonal phase to be about a=3.99 Å, c=7.22 Å, which is slightly larger than the bulk values[10,11,13], possibly due to the composition difference. For Ru(10) and Ru(20) films, similar to Ru(30), they only exhibit the peak of the hexagonal phase at 40.9°, as indicated by the dotted line in Fig. 1(c). The structure of the Ru-30nm sample was further clarified by TEM, as seen in Fig. 2. From the cross-section view, the stack of the films is clearly resolved and the thickness is accurately determined. As the crystal grains of the film are quite small (~10nm), the local electron diffraction pattern in Fig. 2(b) shows a rough polycrystalline pattern. But all the rings can be indexed to the D0$_{19}$ structure, with a matching lattice constant obtained from the XRD.

Temperature-dependent out-of-plane *M-B* loops for the samples of Ta-only, and Ru=10, 30nm are shown in Fig. 3, where the diamagnetic background has been subtracted. In-plane magnetization loops are shown in Fig. S1. There is no distinct magnetic anisotropy is found in our films. The low temperature *M-B* loops for Ta-only and Ru=10nm sample show obvious discontinuous behavior, indicating the coexistence of two phases: one is hard magnetic and the other is soft magnetic or AFM phase. At higher temperatures (above ~100K), the signature of two phases becomes unapparent, caused by the nearly vanished coercivity of the hard phases. But the room temperature XRD of Ta-only sample in Fig. 1(c) still showed the existence of hexagonal and tetragonal phases. We attribute the hard magnetic phase to $D0_{22}$ phase, owing to its reported large magneto-crystalline anisotropy and coercivity[12,26,27]. The AFM phase is taken as the hexagonal $Mn_3Ga$, whose existence is further revealed by the exchange bias (~600–900Oe) presented in Ta(5) and Ta(5)/Ru(10) samples. Moreover, the saturated magnetic moment for Ta(5) film is about 1.2$\mu_B$/f.u. (at 10K) (Fig. 3(a)), approaches the reported ones of tetragonal $Mn_3Ga$[12,13]. It decreases with increasing Ru thickness, becomes only 0.3$\mu_B$/f.u. for Ta(5)/Ru(30) sample, again indicating the transition from the tetragonal phase to nearly antiferromagnetic hexagonal phase. Furthermore, in the Ru=30nm sample, a hysteric behavior appeared in the vicinity of the saturation field, but the no coercivity is present through the whole temperature range, indicating the complete disappearance of tetragonal phase. In some cases, the room temperature coercivity of hexagonal $Mn_3Ga$ film can reach as high as 3T[22]. We attribute this discrepancy to the difference in the crystallization state and film thickness. Take the bulk $Mn_3Ga$ reported in Ref. 9 as an example, the coercivity of the hexagonal phase is also found to be small.

We have further measured the magnetoresistance (MR) and Hall effect in the perpendicular field for the above samples. Representative MR data of Ta/Ru(30) and Ta-only sample are shown in the supplementary Fig. S2. The overall MR value is smaller than 1%, and for Ru(30) sample, it is only 0.03%. Thus the anomalous Hall coefficient $R_A \sim a\rho_{xx} + b\rho_{xx}^2$ can be taken as field independent value when fitting the Hall curve. The Hall resistivity $\rho_{xy}(B)$ curves for various temperatures are shown in Fig. 4. We also calculated the Hall conductivity of them by $\sigma_{xy} = -\rho_{xy}/\rho^2$, where $\rho$ is the normal resistivity, as seen in supplementary Fig. S3. Consistent with the magnetic properties, $\rho_{xy}(B)$ for the Ta-only sample shows a large hysteresis at low temperatures, which originates from the high anisotropy of the tetragonal phase. With the insertion of a 10nm-Ru layer, the coercivity remains but the curve becomes less square due to the existence of antiferromagnetic components.

It is known that for a common ferromagnet, the Hall resistivity can be expressed as $\rho_{xy} = R_oB + R_AM(B)$, where the first and second term represent the normal and anomalous Hall effect, respectively. We applied linear fitting to the high-field Hall curve, as seen in Fig. 5(a). Thus we obtained the value of $R_0$ and $\rho_{xy}^0$ (=$R_AM_S$) from the slope and intercept of the fitting, as shown in Figs. 5(b) and (c). The anomalous Hall resistivities $\rho_{xy}^0$ for the Ru(0) and Ru(10) samples are relatively large owing to their large magnetization. Remarkably, for Ru(30) film, there is nearly no spontaneous anomalous Hall effect despite the non-zero magnetization. The associated Hall conductivity $\sigma_{xy}$ in Fig. S3(c) show a relatively large value of about 60 $\Omega^{-1}\text{cm}^{-1}$ at 5T, but the anomalous Hall term $\sigma_{xy}^A$ (extrapolate the linear fit to zero field) is rather small. This seems inconsistent to the large theoretical values ( 81 $\Omega^{-1}\text{cm}^{-1}$) of D0$_{19}$-type Mn3Ga predicted by Berry curvature calculations[6]. But it should be noted that the large $\sigma_{xy}^A$ can only be observed in specific directions due to the non-coplanar spin configuration[4,5,6]. When the field is aligned along the *c*-axis of Mn$_3$Z, i.e. in the [001] direction, the $\sigma_{xy}^A$ is negligibly small. In polycrystalline Mn$_3$Ga, a moderate $\sigma_{xy}^A$ was observed due to the averaging effect in all directions[9]. Here our Ru(30) Mn$_3$Ga film is textured to be preferably oriented to the *c*-axis ([002] direction), as revealed from the above XRD and TEM results, thus the $\sigma_{xy}^A$ can be rather small. In this reduced Hall conductivity background, a distinguishing hump structure in the low field region is discerned. As seen in Fig. 4(c), the abnormal dome appears when the field is increased from a demagnetization state, but vanishes while decreasing from the saturated field to zero, leading to an asymmetric character. The hump is relatively broad, accompanying with a large field hysteresis in the $\rho_{xy}(B)$ loops.

For multilayer films, the contribution of different layers to the total conduction needs to be considered. We have measured the Hall effect of pure Ru film (not shown), which are linear curves for all temperatures, thus the Hall anomaly can only be attributed to the specific magnetic properties of Mn3Ga film. Moreover, considering the layer thickness for the Ru and Mn$_3$Ga layers (30nm and 200nm, respectively), the surface electronic conduction contributed by the under Ru layer can be minor. Therefore, for simplicity, we use a rough one carrier model to fit the high field $\rho_{xy}(B)$ curve. A similar anti-symmetrical dome structure has also been found in the Hall effect of some magnetic/non-magnetic bilayer heterostructures[28,29], and the interface driven skyrmion formation are taken to be accounted for that. Here due to the existence of Mn-kagome lattice, Mn$_3$Ga has the possibility to form skyrmion domains and give rise to the topological Hall effect.

To investigate the abnormal part of $\rho_{xy}^A$, we subtracted the normal and anomalous Hall term by using the fitted value of R$_o$ and R$_A$, as shown in the representative cure at 100K in Fig. 6(a). As the anomalous Hall term is negligibly small compared with the ordinary one, the additional Hall contribution, here termed as $\rho_{xy}^{ex}$, was extracted and presented in Fig. 6(b) for various temperatures by simply subtracting a linear part. It can be seen that, in a large temperature range from 10 to 100K, $\rho_{xy}^{ex}$ maintains an almost constant value of 8nΩcm, which is comparable with the topological Hall resistivity obtained in skyrmion host MnSi film[15]. When the temperature further increased, $\rho_{xy}^{ex}$ decreased and presented a sign reversal above 200K. The sign reversal is probably caused by some kind of spin reorientation, as the magnetization of the Mn-triangle can be sensitive to the external field[5,30]. Further direct domain observations, such as the Lorentz transmission electron microscope imaging, are necessary to confirm it. In fact, similar sign reversal has also been observed in the topological Hall effect of MnSi film, which is attributed to the uncertainty of spin polarization sign near the Fermi level[15].

The above abnormal Hall effect is reproduced in the Ru(20) buffered films and also in other Ru(30) samples (see Supplementary Fig. S4 and S5). In all these samples, the magnitude of anomalous Hall conductivity can be a bit different, but the hump structures are all clearly observed. One thing we note is that the hexagonal to orthogonal structural transition, which occurred in the bulk polycrystalline $Mn_3Ga$ alloys[9,31], was absent in our films. This is most probably caused by the strain from the substrate, like in normal case of films, the structural transformation are usually suppressed.

## IV. CONCLUSION

In conclusion, by simply tuning the Ta/Ru seedlayer configuration, we acquired $Mn_3Ga$ film in both tetragonal and hexagonal form. The Ta(5)-only samples exhibit mixed-tetragonal $Mn_3Ga$ phase, and present a large coercivity and significant anomalous Hall effect. With increasing Ru thickness, a nearly pure hexagonal $D0_{19}$ $Mn_3Ga$ phase was obtained, accompanying with a nearly vanished coercivity and anomalous Hall resistivity. Remarkably, in Ta(5)/Ru(30) sample, an abnormal hump that resembles the topological Hall effect appeared in the low field region of the Hall curve. The extracted additional Hall term persists in a large temperature range and presents a sign reversal above 200K. The Hall anomaly is supposed to be related to the non-zero spin chirality of the Mn-kagome lattice and indicative of special domain structures like skyrmion domain formation.

## SUPPLEMENTARY MATERIAL

see Supplementary Material for the in-plane magnetization loops for Ta(5) and Ta(5)/Ru(30) sample, and some extra Hall resistivity and conductivity data.

## ACKNOWLEDGEMENTS

This work is sponsored by National Natural Science Foundation of China (11604148, 51571121), Fundamental Research Funds for the Central Universities (30916011344), Natural Science

Foundation of Jiangsu Province (BK20160829, BK20140035), State Key Lab of Advanced Metals and Materials (2015- ZD02). It was also funded by the Qing Lan Project, the Six Talent Peaks Project in Jiangsu Province.

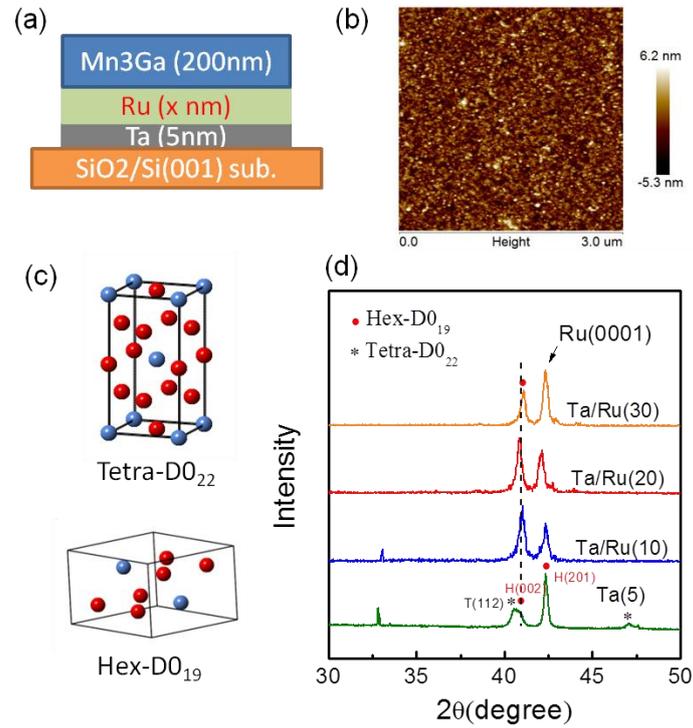

FIG. 1. (a) Stack of the multilayer film, with various Ru thickness. (b) AFM topographical images (scan area 3×3 μm$^2$) of Mn$_3$Ga with Ru-30nm. (c) The crystal structure of tetragonal D0$_{22}$ and hexagonal D0$_{19}$ Mn$_3$Ga. (d) Normalized XRD patterns for Mn$_3$Ga film with different seed layers and the substrate, with the dot and star indicating the peaks for hexagonal D0$_{19}$ and tetragonal D0$_{22}$ structure, respectively.

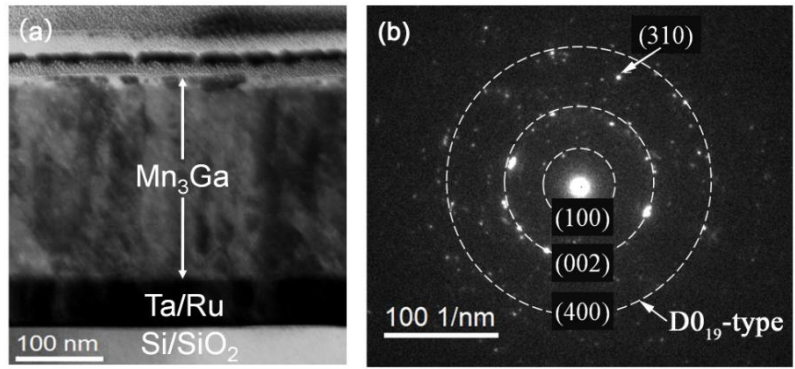

FIG. 2 (a) Cross-section TEM images of the multilayer films and (b) selected area diffraction pattern (SAED) of Ru-30nm sample. The dotted lines are eye-guiding to the diffraction ring, and all the index belongs to the set of $D0_{19}$ structure.

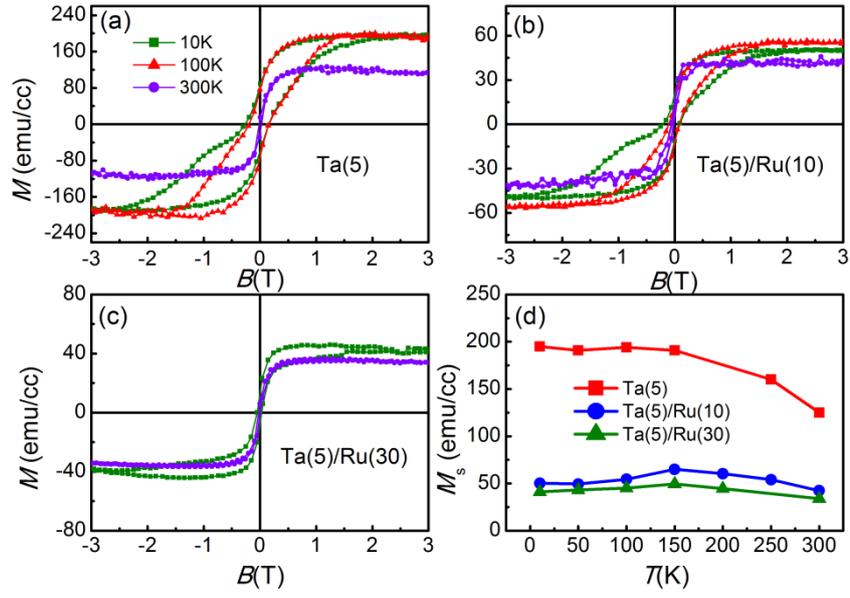

FIG. 3. Out-of-plane hysteresis loops at various temperatures for samples of (a) Ta(5)-only, (b)Ta (5)/Ru(10) and (c)Ta (5)/Ru(30), where the former two ones show a coexistence of tetragonal and hexagonal phase, and the last one shows a nearly pure hexagonal phase. (d) Temperature dependence of the saturated magnetic moment $M_s$ for the three sets of films.

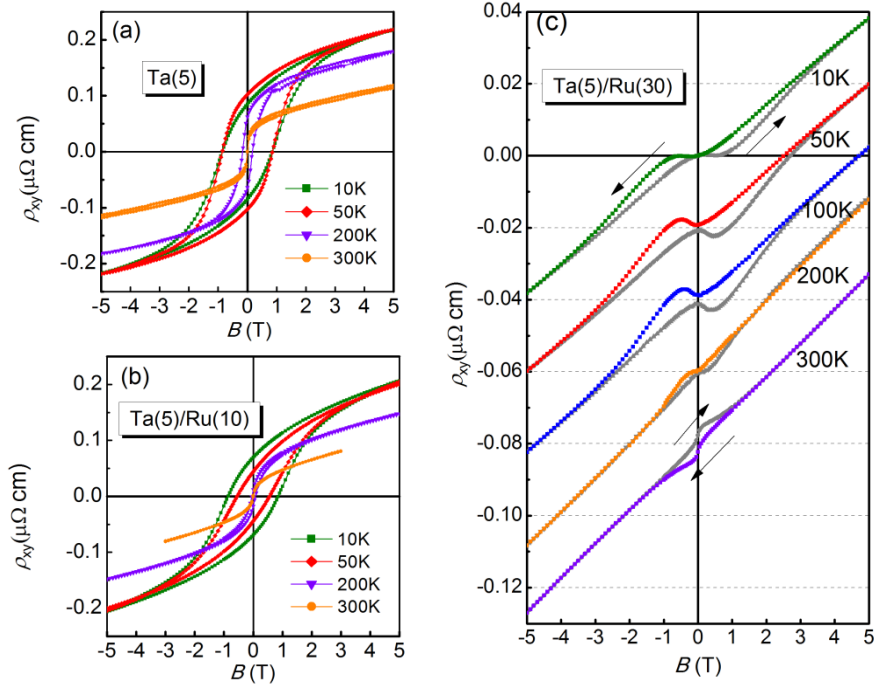

FIG. 4. Magnetic field dependence of the Hall resistivity $\rho_{xy}$ at several temperatures for samples with seed layers of (a) Ta(5)-only, (b) Ta(5)/Ru(10), (c) Ta(5)/Ru(30). Arrows indicated the sweep direction of the magnetic field. In the last panel, the curve was shifted down with increment of 0.02 in the case of higher temperatures. The fitting of ordinary and anomalous Hall coefficient are applied in the high field linear region (~3-5T).

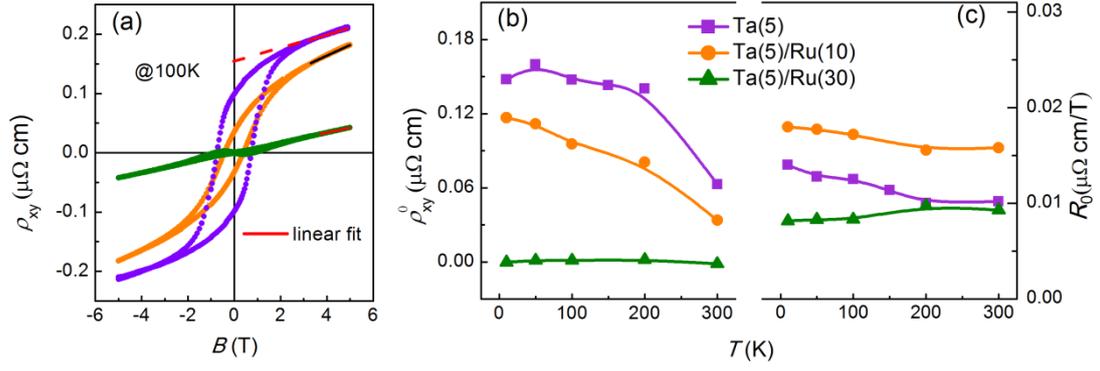

FIG. 5. (a) A comparison of Hall resistivity $\rho_{xy}$ for Ta(5)-only (purple), Ta(5)/Ru(10) (orange), and Ta(5)/Ru(30) (green) buffered Mn$_3$Ga film, and their high field fitting curve (line). (b) and (c) are the fitted zero field anomalous Hall resistivity $\rho_{xy}^0$ ordinary Hall coefficient R$_0$ for various temperatures, respectively.

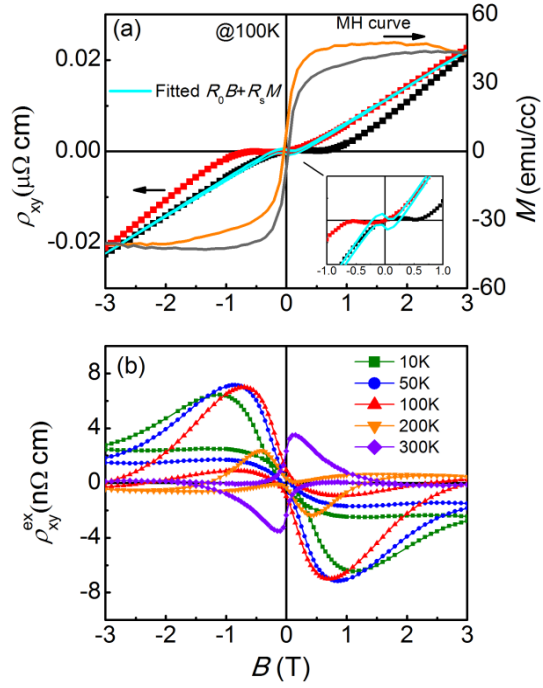

FIG. 6 (a) Typical fitting of the anomalous Hall effect with $R_o B + R_S M(B)$ (cyan line). Inset is the magnified view of original and fitted Hall data in the low field region. (b) The additional Hall curve $\rho_{xy}^{ex}(B)$ at various temperatures for Ru (30) sample, after subtracting of the above linear term.